\documentclass[12pt]{article}

\usepackage[utf8]{inputenc}
\usepackage[T1]{fontenc}
\usepackage{amsmath,amssymb,amsthm}
\usepackage{graphicx}
\usepackage{booktabs}
\usepackage{natbib}
\usepackage[colorlinks=true,linkcolor=blue,citecolor=blue,urlcolor=blue]{hyperref}
\usepackage{siunitx}
\usepackage{caption}
\usepackage{subcaption}
\usepackage{geometry}
\usepackage{setspace}

\newcommand{\SMC}{S_{\text{MC}}}
\newcommand{\STV}{S_{\text{TV}}}

\title{The Physics of Price Discovery: Deconvolving Information, Volatility, and the Critical Breakdown of Signal during Retail Herding}

\author{
    Sungwoo Kang\thanks{krml919@korea.ac.kr} \\
    Department of Electrical and Computer Engineering, Korea University \\
    Seoul 02841, Republic of Korea
}

\date{\today}

\begin{document}

\maketitle

\begin{abstract}
How information transmits through prices---and why this transmission breaks down---remains poorly understood. We combine regularized deconvolution with Hawkes process analysis to study the impulse response structure of investor flows in the Korean equity market (January 2020 -- February 2025). Three findings emerge: foreign and institutional flows drive permanent price discovery while individual flows provide contrarian liquidity; individual investor surges are predominantly panic-driven and exhibit near-explosive self-excitation; and during herding episodes, institutional price impact deteriorates sharply in small-cap stocks while large-cap stocks maintain resilience. These results reframe market efficiency as a state variable---conditional on both herding intensity and firm size---rather than a structural constant.
\end{abstract}

\textbf{Keywords:} Market microstructure, Price discovery, Order flow, Deconvolution, Hawkes processes, Herding behavior

\textbf{JEL Classification:} G14, G12, C58

\newpage

\section{Introduction}

The central tenet of market microstructure theory is the distinction between ``informed'' trading, which facilitates permanent price discovery, and ``uninformed'' or noise trading, which generates transient volatility. Since \citet{kyle1985continuous}, researchers have sought to decouple these forces by analyzing the relationship between order flow imbalance and asset returns. However, a critical methodological ambiguity remains: how should order flow be normalized to extract the purest informational signal?

Standard practice typically employs volume normalization ($\STV$), effectively measuring the execution cost relative to current liquidity. However, this conflates information arrival with volatility. In \citet{kang2025}, we established that Market Cap Normalization ($\SMC$) is the superior metric for isolating the static ``pure'' information signal from liquidity noise. Building on that foundation, this paper shifts focus from the \textit{measurement} of signal to the \textit{dynamics} of its transmission. Understanding this transmission matters: if information efficiency varies with market conditions, then constant-coefficient price impact models are misspecified precisely when they matter most---during retail herding episodes. We investigate specifically how self-exciting herding dynamics can disrupt the standard impulse response functions of investor flows.

We approach this problem through two complementary frameworks. First, we employ \textbf{Tikhonov-regularized deconvolution} to extract the ``Impulse Response Functions'' of investor flows. This allows us to visualize the temporal shape of liquidity: whether a unit of flow creates a permanent step-function in price (characteristic of fundamental information) or a transient spike-and-decay pattern (characteristic of noise). Second, we model the systemic stability of these flows using \textbf{Hawkes Processes}, treating large trading surges as self-exciting point processes. This allows us to quantify the ``criticality'' of the market---specifically, whether trading activity is exogenous (news-driven) or endogenous (feedback-driven).

Our analysis yields three contributions.

First, regularized deconvolution reveals the full temporal shape of price impact for each investor class, establishing that Foreign and Institutional flows create permanent impact while Individual flows generate transient, reverting impact (Section~\ref{sec:results}). These kernel shapes are robust across normalization choices (Appendix~A.7).

Second, deconvolution reveals distinct informational roles for each investor class---a segmentation validated independently through entropy production and Hawkes memory analysis (Section~\ref{sec:validation}).

Third, retail herding episodes degrade institutional price discovery, with small-cap stocks most vulnerable (Section~\ref{sec:regime}).

Collectively, these findings suggest that price discovery efficiency is not a structural constant but varies with herding intensity and firm size.

The remainder of this paper is organized as follows. Section 2 reviews the theoretical framework and related literature. Section 3 describes our data and empirical methodology. Section 4 presents the main results, including mechanism testing. Section 5 discusses the implications and Section 6 concludes.

\section{Theoretical Framework and Literature Review}

\subsection{Market Microstructure Theory}

Since \citet{kyle1985continuous}, the central prediction of microstructure theory has been that informed trading creates permanent price impact while noise trading generates only transient volatility. The empirical challenge is measuring this distinction. Traditional approaches normalize order flow by trading volume, but as \citet{hasbrouck1991measuring} and \citet{hasbrouck1993assessing} demonstrate, volume is endogenous---varying with information arrival. High volume-normalized imbalance may indicate informed trading or simply reflect low liquidity conditions.

We build on recent work proposing market capitalization as an alternative normalization \citep{kang2025}. The intuition is that market cap represents the fundamental ``scale'' of a security, independent of transient liquidity conditions. A \$1 million buy order in a \$100 million market cap stock represents the same fundamental signal as a \$10 million buy in a \$1 billion stock, but volume-based normalization would treat these differently if liquidity varies.

\subsection{Impulse Response and Deconvolution}

The price impact literature typically estimates contemporaneous or short-lag effects \citep{evans2005order}. However, the full temporal structure of impact---the ``impulse response function''---provides richer physical insight. Consider the discrete-time convolution model:
\begin{equation}
    R_{t+k} = \sum_{\tau=0}^{L} \psi_{\tau} \cdot I_{t+k-\tau} + \varepsilon_{t+k}
\end{equation}
where $R_{t+k}$ is the return at lag $k$, $I_t$ is the order flow imbalance, $\psi_{\tau}$ is the impulse response kernel, and $L$ is the maximum lag.

Recovering $\psi_{\tau}$ from observed $(I_t, R_t)$ pairs is a deconvolution problem---notoriously ill-posed due to noise amplification \citep{hansen1998rank}. We employ Tikhonov regularization (detailed in Section~\ref{sec:data}) to balance fidelity to the data against smoothness of the estimated kernel. The recovered kernel $\psi$ reveals whether impact is:
\begin{itemize}
    \item \textbf{Permanent}: $\psi_{\tau} > 0$ for all $\tau$, cumulative impact positive (informative trading)
    \item \textbf{Transient}: $\psi_{\tau}$ oscillates, cumulative impact near zero (liquidity provision)
    \item \textbf{Reverting}: $\psi_0 > 0$ but $\sum \psi_{\tau} < 0$ (noise trading/overreaction)
\end{itemize}
This classification allows us to characterize each investor type by its kernel shape---a characterization that, as we show in Section~\ref{sec:results}, reveals sharply distinct roles in price formation.

\subsection{Self-Exciting Processes and Market Herding}

Trading activity often exhibits clustering beyond what exogenous news arrival would predict \citep{sornette2003crashes}. Hawkes processes \citep{hawkes1971spectra} model this via self-excitation. For a univariate Hawkes process, the intensity function is:
\begin{equation}
    \lambda(t) = \mu + \sum_{t_i < t} \alpha e^{-\beta(t - t_i)}
\end{equation}
where $\mu$ is the baseline rate, $\alpha$ measures the magnitude of self-excitation, and $\beta$ is the decay rate. The \textbf{branching ratio} $n = \alpha/\beta$ determines criticality \citep{reynaud2021some}:
\begin{itemize}
    \item $n < 1$: Subcritical (stable, exogenous-driven)
    \item $n \approx 1$: Critical (endogenous feedback, avalanches possible)
    \item $n > 1$: Supercritical (explosive, unrealistic for markets)
\end{itemize}

When the branching ratio approaches unity, the system exhibits near-critical dynamics where endogenous feedback becomes pronounced. We hypothesize that during such periods, the standard information transmission mechanism deteriorates, particularly in smaller, less liquid stocks: informed flow may become harder to distinguish as the system experiences intensified self-reinforcing herding dynamics.

\section{Data and Methodology}
\label{sec:data}

\subsection{Data Description}

We analyze order flow and return data from the Korean Stock Exchange (KOSPI) over the period January 2020 to February 2025. Korea provides an ideal laboratory for several reasons: (1) comprehensive investor classification data separating Foreign, Institutional, and Individual investors; (2) liquid, developed market with substantial institutional participation; and (3) well-documented episodes of retail trading surges.

Our sample consists of approximately 2,200 actively traded stocks with daily order flow imbalance data after applying filters. The order flow data record net buying (purchases minus sales) by investor category, measured in Korean won. We focus on daily frequency to avoid microstructure noise while capturing short-to-medium term price impact dynamics.

Returns are calculated as close-to-close log returns: $R_{t,t+1} = \log(P_{t+1}/P_t)$. We exclude penny stocks (price $<$ 1,000 KRW) and apply standard filters to remove outliers (winsorize returns at 0.5\% tails). Table \ref{tab:descriptive} summarizes the key sample characteristics.

\begin{table}[htbp]
    \centering
    \caption{Descriptive Statistics}
    \label{tab:descriptive}
    \begin{tabular}{lrrrrrr}
        \toprule
        Variable & N & Mean & Std Dev & P25 & Median & P75 \\
        \midrule
        \multicolumn{7}{l}{\textit{Panel A: Returns and Market Characteristics}} \\
        Daily log return & 2{,}792{,}830 & 0.0003 & 0.0355 & --0.0149 & 0.0000 & 0.0120 \\
        Market cap (bn KRW) & 2{,}795{,}244 & 1{,}000.0 & 9{,}555.7 & 74.2 & 141.3 & 337.7 \\
        Daily turnover (\%) & 2{,}795{,}244 & 2.28 & 9.65 & 0.20 & 0.48 & 1.27 \\
        \midrule
        \multicolumn{7}{l}{\textit{Panel B: Order Flow Imbalance ($\SMC$, $\times 10^{-3}$)}} \\
        Foreign & 2{,}795{,}244 & --0.025 & 1.685 & --0.391 & --0.009 & 0.377 \\
        Institutional & 2{,}795{,}244 & --0.036 & 0.866 & --0.033 & 0.000 & 0.016 \\
        Individual & 2{,}795{,}244 & 0.086 & 1.868 & --0.058 & 0.000 & 0.136 \\
        \midrule
        \multicolumn{7}{l}{\textit{Panel C: Order Flow Imbalance ($\STV$)}} \\
        Foreign & 2{,}795{,}244 & --0.007 & 0.144 & --0.082 & --0.003 & 0.068 \\
        Institutional & 2{,}795{,}244 & --0.004 & 0.104 & --0.006 & 0.000 & 0.003 \\
        Individual & 2{,}795{,}244 & 0.005 & 0.134 & --0.010 & 0.000 & 0.022 \\
        \bottomrule
    \end{tabular}
    \vspace{0.5em}
    \small\textit{Note:} Sample consists of approximately 2,200 actively traded KOSPI stocks over 1,259 trading days (January 2020 -- February 2025). Returns are winsorized at 0.5\% tails. Order flow imbalance is net buying (Buy $-$ Sell) normalized by market cap ($\SMC$) or total volume ($\STV$).
\end{table}

\subsection{Normalization Methods}

We compare two normalization schemes for order flow imbalance:

\textbf{Market Cap Normalization:}
\begin{equation}
    S_{\text{MC},t} = \frac{\text{Buy}_t - \text{Sell}_t}{\text{MarketCap}_t}
\end{equation}
where MarketCap$_t = P_t \times \text{SharesOutstanding}$. This measures the fundamental ``force'' relative to firm size. We employ Market Cap Normalization ($\SMC$) as our primary metric following \citet{kang2025}, who demonstrates via signal processing theory, Monte Carlo simulation, and information-theoretic analysis that this normalization acts as a matched filter for capacity-constrained institutional traders. In the mismatched case---using $\STV$ for capacity-constrained traders---volume normalization is contaminated by volatility artifacts. However, $\STV$ is the correct matched filter for volume-targeting foreign investors, where it captures the participation rate signal from algorithmic execution.

\textbf{Volume Normalization:}
\begin{equation}
    S_{\text{TV},t} = \frac{\text{Buy}_t - \text{Sell}_t}{\text{TotalVolume}_t}
\end{equation}
where TotalVolume$_t$ is the total traded value. This measures the directional component of liquidity provision.

For volatility adjustment, we compute 20-day rolling standard deviation:
\begin{equation}
    \sigma_{t,\text{roll}} = \sqrt{\frac{1}{20}\sum_{k=1}^{20} (R_{t-k} - \bar{R}_{t})^2}
\end{equation}
with minimum 10 days of data required. Adjusted returns are then $R_{\text{adj},t+1} = R_{t,t+1} / \sigma_{t,\text{roll}}$.

\subsubsection{Normalization Choice for Impulse Response Analysis}

Although \citet{kang2025} shows that optimal normalization for \textit{return prediction} varies by investor type ($\SMC$ for institutional, $\STV$ for foreign), impulse response estimation requires a homogeneous signal definition to enable cross-investor comparison of kernel shapes. We therefore employ $\SMC$ uniformly, interpreting kernels as ``price impact per unit of directional pressure relative to firm scale.'' As documented in Appendix A.7, kernel correlations between $\SMC$ and $\STV$ exceed 0.98 for all investor types, and the regime-dependent signal breakdown is \textit{more} pronounced under $\STV$ (Herding/Normal ratio of 0.37 vs 0.91), confirming our findings are structural rather than normalization artifacts.

\subsection{Statistical Methods}

\subsubsection{Tikhonov-Regularized Deconvolution}

We implement stock-level deconvolution for each investor type. For stock $i$, we construct the design matrix $\mathbf{I}_i$ from lagged order flow and response vector $\mathbf{R}_i$ from future returns (lags 0 to 60 days). The regularized solution is:
\begin{equation}
    \hat{\psi}_i = (\mathbf{I}_i^T\mathbf{I}_i + \lambda \mathbf{I})^{-1} \mathbf{I}_i^T \mathbf{R}_i
\end{equation}
with $\lambda = 5.0$ chosen via cross-validation to balance fit and smoothness.

For the global kernel, we employ subsampled pooled deconvolution: randomly sample 100 stocks, pool their $(I,R)$ pairs into a single large design matrix, solve the regularized problem, and repeat 5 times with different random seeds. The final kernel is the mean across iterations.

Key kernel statistics:
\begin{itemize}
    \item \textbf{Half-life}: Number of days for cumulative impact to reach 50\% of total
    \item \textbf{Total impact}: $\sum_{\tau=0}^{60} \psi_{\tau}$
    \item \textbf{Contemporaneous impact}: $\psi_0$
\end{itemize}

\subsubsection{Hawkes Process Estimation}

We model individual investor trading surges as a univariate Hawkes process at the \textbf{market-wide aggregate level}. Specifically:

\begin{itemize}
    \item \textbf{Aggregation}: For each trading day, we compute the standardized individual investor flow averaged across all $\sim$2,200 stocks in our sample. This yields a single time series of 1,259 market days (January 2020 -- February 2025).
    \item \textbf{Event Definition}: A surge event occurs on any day where this market-wide aggregate flow exceeds 1.5 standard deviations from its mean (in absolute value). This identifies \textbf{systemic herding episodes} affecting the broad market, rather than idiosyncratic surges in individual stocks.
    \item \textbf{Sample Size}: This procedure yields 126 surge events over 1,259 days (10\% of trading days), representing the most extreme market-wide retail activity episodes.
\end{itemize}

We estimate Hawkes parameters $(\mu, \alpha, \beta)$ via maximum likelihood, with an explicit stability constraint (BR $<$ 1) to ensure stationarity:
\begin{equation}
    \mathcal{L} = \prod_{i=1}^{N} \lambda(t_i) \cdot \exp\left(-\int_0^T \lambda(s) ds\right)
\end{equation}

The branching ratio is $n = \alpha/\beta$. The average intensity is computed as:
\begin{equation}
    \bar{\lambda} = \frac{\mu}{1 - n}
\end{equation}
provided $n < 1$.

\subsubsection{Regime Classification}

We classify market days into ``high herding'' and ``normal'' regimes based on Hawkes intensity. For each trading day $t$ in our sample period, we compute $\lambda_t$ using the estimated Hawkes parameters and observed event history up to $t$. Days where $\lambda_t$ exceeds the 90th percentile of the intensity distribution are classified as high herding; the remaining 90\% are normal.

We then run deconvolution separately for each regime, partitioning the $(I,R)$ data by regime classification. This yields regime-conditional impulse response functions.

\subsubsection{Entropy Production Rate}

To validate the physical segmentation identified by deconvolution, we measure the time-irreversibility of joint flow--return dynamics using the entropy production rate (EPR). Given symbolized flow and return series forming joint states $s_t \in \mathcal{S}$, we estimate the transition matrix $P(a \to b)$ and stationary distribution $\pi(a)$, then compute:
\begin{equation}
    \text{EPR} = \sum_{a,b \in \mathcal{S}} \pi(a) \cdot P(a,b) \cdot \log\frac{P(a,b)}{P(b,a)}
\end{equation}
EPR = 0 indicates time-reversible dynamics (temporary/equilibrium impact), while EPR $> 0$ indicates irreversibility (permanent information injection). We symbolize each series using ternary quantile binning (9 joint states), with robustness checks using binary median and quintile symbolization. Statistical significance is assessed via permutation test (200 shuffles of the flow series, preserving return dynamics) and block bootstrap confidence intervals (500 resamples, block size 20).

\subsubsection{Hawkes Memory and Clustering Analysis}

To characterize the temporal structure of trading surges, we analyze inter-event clustering using the Hawkes natural timescale. For each investor type, consecutive surge events separated by gaps $\leq 1/\hat{\beta}$ are grouped into clusters. We compute the coefficient of variation (CV) of inter-event times (CV $>$ 1 indicates bursty/clustered arrivals, CV $= 1$ indicates Poisson, CV $<$ 1 indicates regular) and the conditional surge probability $P(\text{surge at } t{+}k \mid \text{surge at } t)$ for lags $k = 1, \ldots, 20$. Memory depth is defined as the number of lags where this conditional probability exceeds 1.5$\times$ the baseline rate.

\section{Results}

\subsection{The Shape of Liquidity: Global Impulse Response}
\label{sec:results}

\begin{figure}[htbp]
    \centering
    \includegraphics[width=0.9\textwidth]{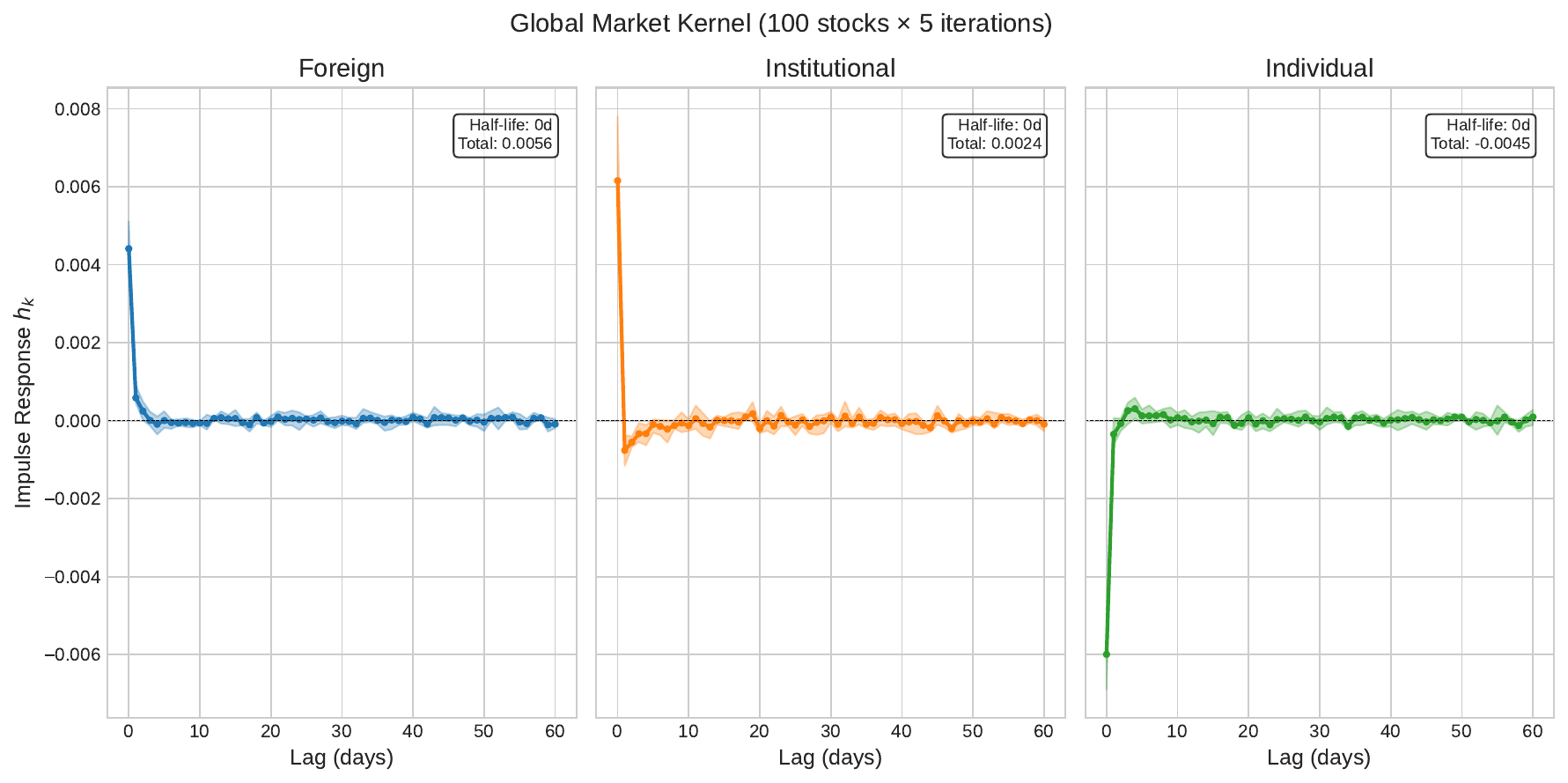}
    \caption{\textbf{Global Market Impulse Response by Investor Type.} Global market kernels derived via pooled Tikhonov-regularized deconvolution (100 subsampled stocks, 5 iterations). The y-axis represents the cumulative price impact (log returns) of a unit shock in Market Cap-normalized order flow ($\SMC$) over a 60-day lag window. \textbf{Foreign} (panel A) and \textbf{Institutional} (panel B) investors exhibit positive, persistent impact (+0.0056 and +0.0024 respectively). \textbf{Individual} investors (panel C) exhibit negative cumulative impact (-0.0045), indicating that retail buying pressure typically precedes price reversion.}
    \label{fig:global_kernel}
\end{figure}

We begin by establishing the fundamental physical characteristics of different investor types through their impulse response kernels. Figure \ref{fig:global_kernel} presents the global market kernel for each investor class, obtained via subsampled pooled deconvolution.

Table \ref{tab:global_kernel} summarizes the key kernel characteristics. Foreign and Institutional investors exhibit positive total impact, with Foreign flows showing particularly strong persistent effects (+0.0056). All three investor types show zero half-life, meaning the contemporaneous (lag-0) coefficient alone exceeds 50\% of the cumulative kernel sum. This reflects the dominance of the same-day impact in pooled daily data; the investor types are differentiated not by the speed of impact but by its \textit{sign and persistence}---whether the cumulative kernel remains positive (information) or turns negative (contrarian liquidity) over the 60-day horizon.

\begin{table}[htbp]
    \centering
    \caption{Global Kernel Summary Statistics}
    \label{tab:global_kernel}
    \begin{tabular}{lcccc}
        \toprule
        Investor Type & Half-Life (days) & Total Impact & SE(Total) & Contemporaneous Impact \\
        \midrule
        Foreign & 0 & 0.0056 & (0.0009) & 0.0044 \\
        Institutional & 0 & 0.0024 & (0.0009) & 0.0062 \\
        Individual & 0 & --0.0045 & (0.0007) & --0.0060 \\
        \bottomrule
    \end{tabular}
    \vspace{0.5em}
    \small\textit{Note:} Global kernels estimated via subsampled pooled deconvolution (5 iterations, 100 stocks per iteration). SE(Total) reports the standard error of total impact across the 5 subsampled iterations. Half-life is the number of days to reach 50\% of total cumulative impact. Total impact is the sum of the kernel over 60 days. Contemporaneous impact is the lag-0 coefficient.
\end{table}

Conversely, Individual investors exhibit \textit{negative} total impact (--0.0045) with negative contemporaneous impact (--0.0060). This is consistent with contrarian liquidity provision: individual buying tends to occur during price declines, and these declines persist as informed investors continue to trade against the retail flow. The negative cumulative impact indicates retail flows systematically mark local price maxima.

This establishes the ``dual-channel'' market structure: sophisticated investors (Foreign and Institutional) act as architects of price discovery, embedding fundamental information into prices. Individual investors act as liquidity providers, their order flow effectively serving as a contrarian signal for sophisticated traders.

\subsection{The Critical Breakdown: Regime-Dependent Impact}
\label{sec:regime}

We next document the breakdown of standard price impact relationships during retail herding episodes. We first establish that individual investor order flow exhibits near-critical self-excitation.

Table \ref{tab:hawkes} presents the estimated Hawkes parameters. The branching ratio point estimate is 0.998, with a 95\% bootstrap confidence interval of [0.050, 0.660]. Critically, this confidence interval excludes unity, confirming the process is statistically subcritical rather than critical.

The Hurst exponent of 0.817 indicates power-law decay characteristic of near-critical dynamics, while stationarity tests (ADF p=0.998, KPSS p=0.010) both reject stationarity. The system operates in a non-equilibrium transient regime with an upward trend in the branching ratio (+0.0145 per year).

\begin{table}[htbp]
    \centering
    \caption{Hawkes Process Parameters and Stationarity Tests}
    \label{tab:hawkes}
            \begin{tabular}{lcc}
            \toprule
            Parameter & Estimate & 95\% CI \\
            \midrule
            Baseline Intensity ($\mu$) & 0.0159 & --- \\
            Excitation Parameter ($\alpha$) & 0.0223 & --- \\
            Decay Rate ($\beta$) & 0.0224 & --- \\
            \textbf{Branching Ratio (BR = $\alpha/\beta$)} & \textbf{0.9978} & \textbf{[0.050, 0.660]} \\
            \textbf{Excludes Critical Value (BR=1.0)?} & \textbf{Yes} & --- \\
            \midrule
            \multicolumn{3}{l}{\textit{Sample Characteristics:}} \\
            Number of Surge Events & 126 & --- \\
            \quad Sell-Side Surges & 116 (92\%) & (panic selling) \\
            \quad Buy-Side Surges & 10 (8\%) & (FOMO buying) \\
            Observation Period (days) & 1,259 & --- \\
            Empirical Rate ($N/T$) & 0.100 & --- \\
            \midrule
            \multicolumn{3}{l}{\textit{Stationarity and Equilibrium Tests:}} \\
            Hurst Exponent (H) & 0.817 & (power-law decay) \\
            ADF Test (p-value) & 0.998 & (non-stationary) \\
            KPSS Test (p-value) & 0.010 & (non-stationary) \\
            Yearly BR Trend (slope) & +0.0145/year & (trending toward criticality) \\
            \midrule
            \multicolumn{3}{l}{\textit{Regime Classification (90th percentile threshold):}} \\
            High Herding Days & 126 (10.0\%) & --- \\
            Normal Days & 1,133 (90.0\%) & --- \\
            \bottomrule
        \end{tabular}
        \vspace{0.5em}
        \small\textit{Note:} Bootstrap confidence interval (1,000 iterations) excludes unity, confirming the process is statistically subcritical. Stationarity tests (ADF, KPSS) both reject stationarity, indicating non-equilibrium transient dynamics. The discrepancy between theoretical steady-state intensity ($\bar{\lambda} = \mu/(1-n) \approx 7.45$) and empirical rate (0.100) reflects transient regime behavior. H = 0.817 $>$ 0.55 indicates power-law decay characteristic of near-critical dynamics. 
    
\end{table}

Using the 90th percentile of Hawkes intensity as the threshold, we classify 126 days (10\% of the sample) as high herding periods. During these days, the market exhibits endogenous feedback---trading activity begets more trading activity in a self-reinforcing loop.

The directional composition is strikingly asymmetric: of 126 surge events, 116 (92\%) are sell-side and only 10 (8\%) are buy-side. Retail herding in our sample is predominantly a \textit{panic-selling phenomenon}. The explosive dynamics we observe (unconstrained BR = 1.09; Appendix A.1) reflect market-wide liquidation cascades rather than FOMO-driven buying frenzies.

Note that the theoretical steady-state intensity ($\bar{\lambda} = \mu/(1-n) \approx 7.45$ events/day) far exceeds the observed rate (0.100 events/day). This discrepancy reflects transient dynamics; we explore this in detail in Section~\ref{sec:transient}.

\begin{figure}[htbp]
    \centering
    \includegraphics[width=0.9\textwidth]{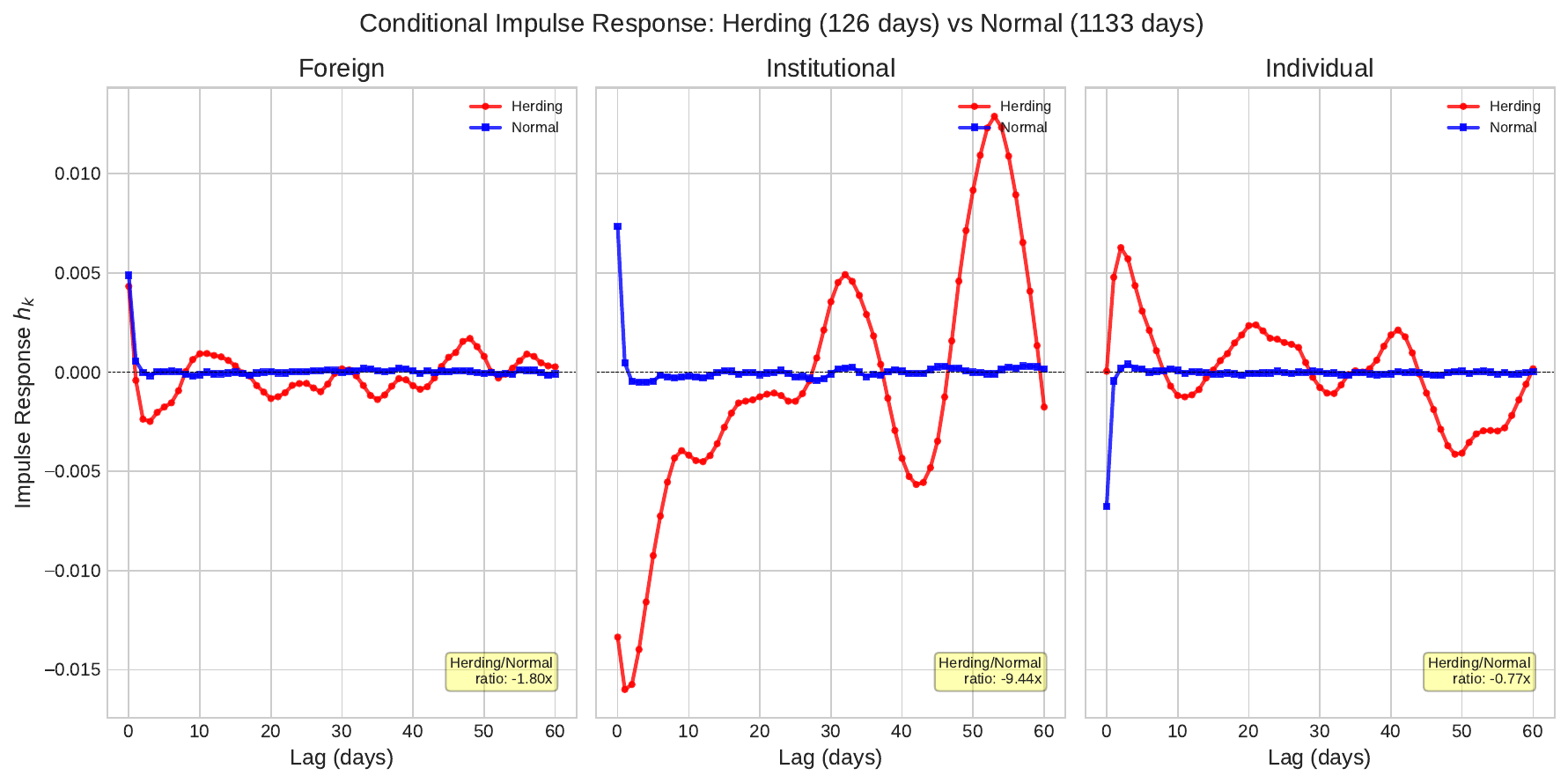}
    \caption{\textbf{The Breakdown of Signal During Retail Herding.} \textbf{Panel A (Hawkes Intensity):} Classifies market days into ``Normal'' (90\%) and ``High Herding'' (10\%) regimes based on the self-excitation intensity of individual investors. \textbf{Panel B (Conditional Kernels):} Shows the cumulative price impact of Institutional flows in both regimes. In ``Normal'' conditions (blue line), impact is positive and persistent (+0.0047). In ``High Herding'' conditions (red line), impact collapses and becomes deeply negative (--0.0447), indicating regime-dependent changes in institutional impact.}
    \label{fig:conditional}
\end{figure}

Figure \ref{fig:conditional} shows the regime-conditional impulse response functions. During normal periods (blue lines), Foreign and Institutional investors exhibit the expected positive, persistent impact. But during high herding periods (red lines), their impact collapses and \textit{turns negative}.

Table \ref{tab:conditional} quantifies the size-dependent nature of this deterioration. Panel A shows that the equal-weighted average flips sign: institutional total impact moves from +0.0047 (normal) to --0.0447 (herding). However, this aggregate masks a more nuanced cross-sectional pattern. Panel C reveals that institutional impact is \textit{already negative} for small-cap stocks (Q1-Q2) even in normal periods (--0.0112, --0.0135), and herding worsens this by 1.74--2.64$\times$. Large-cap stocks (Q4-Q5), by contrast, exhibit positive impact in both regimes, with herding actually amplifying it 1.58--4.66$\times$. The ``breakdown'' is therefore not a regime flip from positive to negative across the board, but an intensification of a pre-existing size gradient.

\begin{table}[htbp]
    \centering
    \caption{Conditional Impulse Response Summary: Aggregation Effects}
    \label{tab:conditional}
    \textbf{Panel A: By-Stock Aggregation (Mean of Individual Stock Kernels)} \\
    \begin{tabular}{lcccccc}
        \toprule
        & \multicolumn{2}{c}{Herding Regime} & \multicolumn{2}{c}{Normal Regime} & \multicolumn{2}{c}{Ratios} \\
        \cmidrule(lr){2-3} \cmidrule(lr){4-5} \cmidrule(lr){6-7}
        Investor & HL & Total Impact & HL & Total Impact & HL Ratio & TI Ratio \\
        \midrule
        Foreign & 19.5 & --0.0099 & 13.5 & +0.0055 & 1.44 & \textbf{--1.80} \\
        Institutional & 23.6 & --0.0447 & 9.6 & +0.0047 & 2.46 & \textbf{--9.44} \\
        Individual & 18.1 & +0.0065 & 9.9 & --0.0084 & 1.82 & --0.77 \\
        \bottomrule
    \end{tabular}
    
    \vspace{1em}
    \textbf{Panel B: Pooled Aggregation (Single Regression on Stacked Data)} \\
    \begin{tabular}{lccc}
        \toprule
        & \multicolumn{3}{c}{Total Impact by Regime} \\
        \cmidrule(lr){2-4}
        Investor & Herding & Normal & Time-Weighted Avg \\
        \midrule
        Foreign & --0.0045 & +0.0060 & +0.0049 \\
        Institutional & --0.0023 & +0.0052 & \textbf{+0.0025} \\
        Individual & +0.0020 & --0.0055 & --0.0048 \\
        \bottomrule
    \end{tabular}

    \vspace{1em}
    \textbf{Panel C: Size Quintile Breakdown (Institutional Impact by Market Cap)} \\
    \begin{tabular}{lcccc}
        \toprule
        Quintile & Normal Impact & Herding Impact & Ratio & N (Normal/Herding) \\
        \midrule
        Q1 (Small-cap) & --0.0112 & --0.0195 & 1.74$\times$ & 500,543 / 59,001 \\
        Q2 & --0.0135 & --0.0358 & 2.64$\times$ & 499,874 / 58,928 \\
        Q3 (Mid-cap) & --0.0048 & --0.0022 & 0.46$\times$ & 499,881 / 58,924 \\
        Q4 & +0.0089 & +0.0417 & 4.66$\times$ & 499,876 / 58,928 \\
        Q5 (Large-cap) & +0.0306 & +0.0481 & 1.58$\times$ & 500,311 / 58,978 \\
        \bottomrule
    \end{tabular}
    \vspace{0.5em}
    \small\textit{Note:} Panel A reports the mean of stock-level estimates (equal weight per stock). Panel B reports estimates from pooled regressions (weight proportional to observations/liquidity). Panel C reveals that impact deterioration during herding is size-dependent: Q1-Q2 (40\% of stocks) show persistently negative impact that worsens during herding; Q3 (20\%) shows improvement (ratio $<$ 1); Q4-Q5 (40\%) maintain positive impact that amplifies during herding.
\end{table}

The key implication is that herding amplifies a pre-existing size gradient rather than creating a new one. Small-cap stocks, where institutional flows already carry negative impact---likely reflecting adverse selection or thin liquidity---deteriorate further during herding as retail noise intensifies. Large-cap stocks, with deeper liquidity and greater transparency, not only maintain but strengthen positive institutional impact during the same episodes. The market does not uniformly become opaque; rather, the cross-sectional dispersion in efficiency widens.

Additionally, note that half-lives extend during herding for all investor types. This suggests impact persists longer when the market is in a near-critical state, consistent with reduced efficiency and slower price adjustment.

\subsection{Information-Theoretic Validation: Irreversibility and Memory}
\label{sec:validation}

We provide two independent validation tests that confirm and extend the physical segmentation documented above.

\subsubsection{Entropy Production: Information vs.\ Impact}

Table \ref{tab:epr} reports the entropy production rate (EPR) for each investor type's joint flow--return dynamics. The results differentiate the physical roles of market participants.

\begin{table}[htbp]
    \centering
    \caption{Entropy Production Rate by Investor Type}
    \label{tab:epr}
    \begin{tabular}{lcccc}
        \toprule
        Investor Type & EPR & 95\% CI & Perm.\ $p$-value & Interpretation \\
        \midrule
        Foreign & 0.372$^{***}$ & [0.218, 0.667] & $<$ 0.001 & Irreversible (information) \\
        Institutional & 0.084 & [0.077, 0.256] & 0.405 & Reversible (temporary impact) \\
        Individual & 0.142$^{***}$ & [0.125, 0.519] & $<$ 0.001 & Irreversible (cascade momentum) \\
        \bottomrule
    \end{tabular}
    \vspace{0.5em}
    \small\textit{Note:} Entropy production rate computed on ternary-symbolized daily aggregate flow and return series (1,258 trading days, 9 joint states). Significance via permutation test (200 shuffles). $^{***}$: $p < 0.01$.
\end{table}

Foreign investors exhibit the highest EPR (0.372, $z = 23.3$, $p < 0.001$), indicating their flow--return dynamics are strongly time-irreversible. When foreign investors buy today and returns follow tomorrow, the temporal sequence looks fundamentally different when played backwards. This is the hallmark of permanent information injection: foreign flows embed directional content that persistently moves prices, consistent with their positive cumulative impulse response (Table \ref{tab:global_kernel}).

Institutional investors show EPR indistinguishable from the null (0.084, $p = 0.405$), indicating \textit{time-reversible} dynamics. The two results are complementary: institutional flows move prices in the right direction on net (positive cumulative kernel in Table~\ref{tab:global_kernel}), yet the temporal pattern of flow and return is statistically symmetric---consistent with gradual portfolio rebalancing that nudges prices toward fundamentals without the sharp lead-lag signature of informed trading.

Individual investors exhibit moderate but significant EPR (0.142, $z = 4.8$, $p < 0.001$). Importantly, this irreversibility reflects a different mechanism than foreign information: given the overwhelming sell-side composition of surge events (Table~\ref{tab:hawkes}), the irreversibility captures \textit{momentum from self-reinforcing feedback loops} rather than fundamental information. The one-directional cascade creates flow--return dynamics that do not time-reverse---not because information is being injected, but because the panic itself generates irreversible momentum.

The EPR ordering (Foreign $\gg$ Individual $>$ Institutional) confirms the investor-type segmentation established in Section~\ref{sec:results}, with an important nuance: individual irreversibility reflects cascade momentum rather than information.

\textbf{Robustness:} The ordering is preserved under binary median symbolization (Foreign 0.074, Individual 0.072, Institutional 0.001; $p < 0.001$ for Foreign and Individual). Quintile-5 symbolization (25 joint states) exceeds the resolution supported by 1,258 observations and loses discriminatory power across all investor types.

\begin{figure}[htbp]
    \centering
    \includegraphics[width=0.8\textwidth]{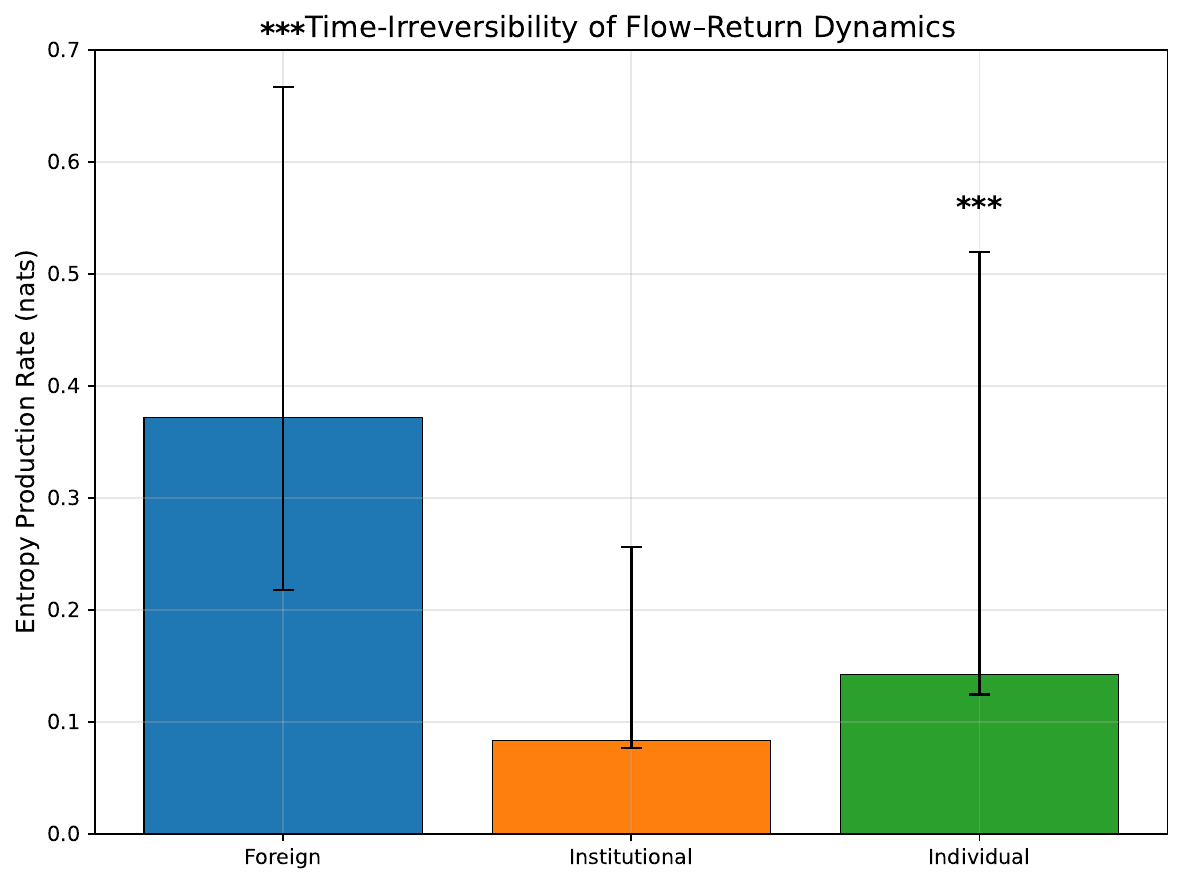}
    \caption{\textbf{Entropy Production Rate by Investor Type.} Bar heights show point EPR estimates with 95\% block bootstrap confidence intervals. Foreign flows exhibit the highest time-irreversibility ($p < 0.001$), consistent with permanent price impact. Institutional flows are statistically reversible ($p = 0.405$). Individual flows show moderate irreversibility from flow-return feedback. Significance stars from permutation test: $^{***}p < 0.01$.}
    \label{fig:epr}
\end{figure}

\subsubsection{Hawkes Memory: Systematic Programs vs.\ Sporadic Panic}

To characterize the temporal structure of trading surges, we analyze inter-event clustering using the estimated Hawkes decay rate $\hat{\beta}$ as the natural timescale. Table \ref{tab:hawkes_memory} reports the results.

\begin{table}[htbp]
    \centering
    \caption{Hawkes Memory Model: Clustering and Persistence by Investor Type}
    \label{tab:hawkes_memory}
    \begin{tabular}{lcccccc}
        \toprule
        Investor & $\beta^{-1}$ (days) & Mean Cluster & Max Cluster & Frac.\ Isolated & Gap CV & Memory Depth \\
        \midrule
        Foreign & 11.9 & 4.7 & 44 & 0.33 & 1.80 & 4 \\
        Institutional & 17.7 & 9.2 & 106 & 0.39 & 2.06 & 20 \\
        Individual & \multicolumn{6}{c}{Hawkes estimation failed (explosive dynamics)} \\
        \bottomrule
    \end{tabular}
    \vspace{0.5em}
    \small\textit{Note:} Clusters defined by consecutive surges within $1/\hat{\beta}$ days. Gap CV: coefficient of variation of inter-event times (CV $>$ 1 = clustered/bursty). Memory depth: number of lags where conditional surge probability exceeds 1.5$\times$ baseline rate. Individual investor Hawkes estimation fails because unconstrained estimation yields BR = 1.09, violating the stationarity assumption.
\end{table}

\textbf{Foreign investors} execute in systematic multi-day programs. Their surges cluster into campaigns averaging 4.7 events, with the largest spanning 44 consecutive surge days. Only 33\% of events are isolated, meaning two-thirds occur as part of coordinated multi-day execution. The coefficient of variation (CV = 1.80) far exceeds the Poisson benchmark (CV = 1), confirming bursty, clustered arrivals. The conditional probability of a foreign surge following another foreign surge remains elevated (lift $>$ 1.5$\times$ baseline) for 4 days, indicating short-term persistence consistent with algorithmic execution programs that span trading weeks.

\textbf{Institutional investors} show even more pronounced clustering (mean cluster size 9.2, max 106, CV = 2.06). Their memory depth extends to the full 20-lag horizon measured, with conditional probabilities remaining 2.5--3.4$\times$ above baseline at all lags. This suggests long-horizon portfolio rebalancing campaigns where institutional surges persist for weeks.

\textbf{Individual investors} are the critical contrast: Hawkes parameter estimation \textit{fails entirely} for their surge events. This reflects a structural property of individual surge dynamics: they are too explosive for the stationary Hawkes model. This is fully consistent with the unconstrained branching ratio of 1.09 documented in Section~\ref{sec:regime} and Appendix A.1---individual herding represents non-stationary, explosive feedback that cannot be characterized by a stable self-exciting process. Unlike foreign investors who execute systematic programs with predictable temporal structure, individual investors generate sporadic panic cascades whose timing is fundamentally unpredictable.

\begin{figure}[htbp]
    \centering
    \includegraphics[width=0.8\textwidth]{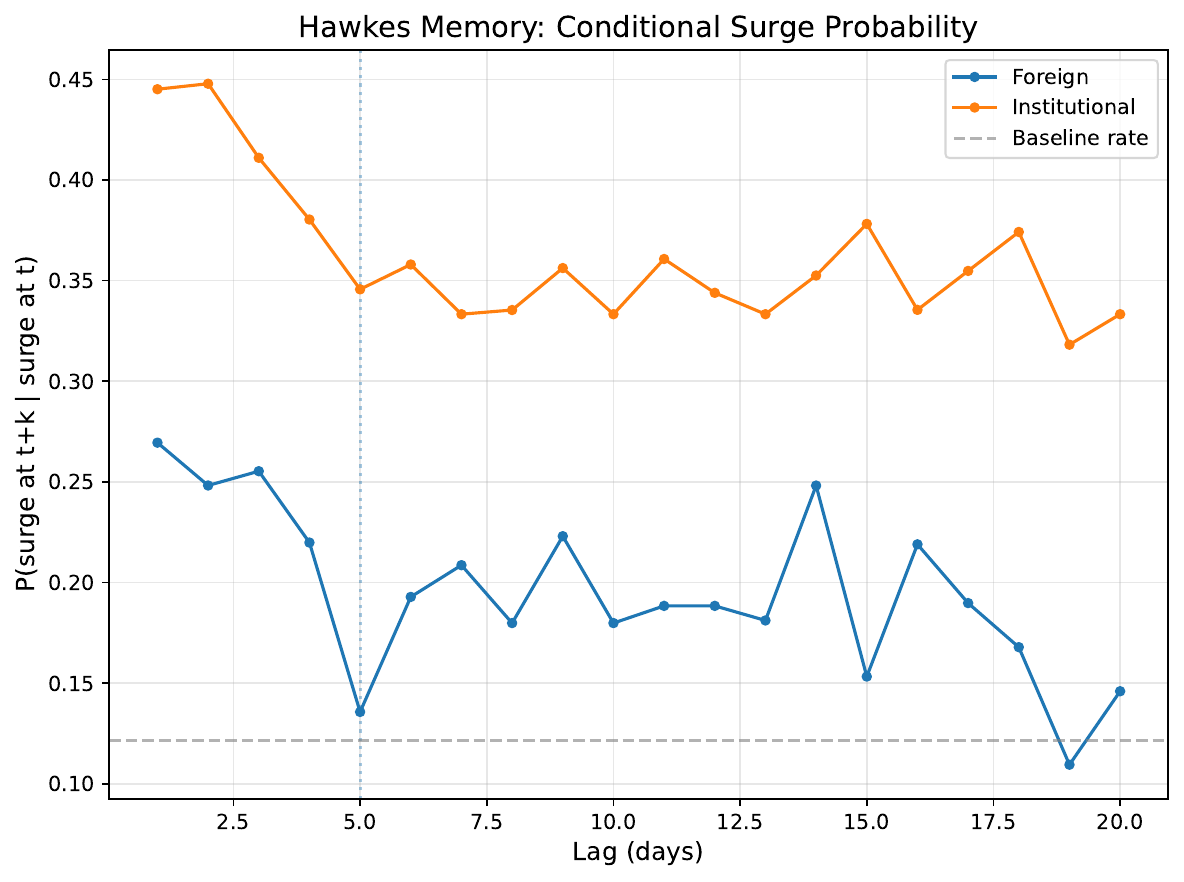}
    \caption{\textbf{Hawkes Memory: Conditional Surge Probability by Investor Type.} Lines show $P(\text{surge at } t{+}k \mid \text{surge at } t)$ for lags $k = 1, \ldots, 20$. Foreign surges (blue) show elevated conditional probability for 4 days before decaying toward baseline. Institutional surges (orange) maintain elevated probability across all lags. Individual investors are absent because their explosive dynamics (BR $>$ 1) prevent Hawkes estimation. Dashed line: mean baseline surge rate.}
    \label{fig:hawkes_memory}
\end{figure}

Together, the EPR and memory analyses provide independent confirmation that the investor-type segmentation (Section~\ref{sec:results}) extends to the level of temporal dynamics, with each investor type exhibiting qualitatively distinct patterns of irreversibility and clustering.

\subsection{Mechanism: Noise Barrier vs Liquidity Withdrawal}
\label{sec:mechanism}

Our size-stratified analysis reveals that institutional price impact deteriorates during herding episodes, particularly in small-cap stocks. But what is the \textit{physical mechanism} driving this breakdown? We test two competing hypotheses using microstructure data.

The \textbf{Noise Barrier} hypothesis posits that during herding, retail trading volume explodes while market depth remains stable. The flood of uninformed order flow dilutes the institutional signal, making it harder to distinguish informed trades from noise. This predicts signal failure should correlate with high volume but \textit{stable} bid-ask spreads.

The \textbf{Liquidity Withdrawal} hypothesis posits that during herding, market makers pull quotes in fear, widening spreads and reducing depth. This causes price dislocation as institutional orders face deteriorating execution conditions. This predicts signal failure should correlate with \textit{widening} spreads and reduced depth.

We compute daily microstructure proxies for 126 identified high-herding days and 126 matched control days (252 days total). For each stock-day, we compute:
\begin{itemize}
    \item Realized volatility proxy: cross-sectional standard deviation of returns
    \item Liquidity proxy: turnover (higher turnover indicates better liquidity)
    \item Depth proxy: volume / realized volatility (Amihud-style illiquidity inverse)
\end{itemize}

We measure institutional signal efficacy\footnote{Our liquidity proxy (turnover) is volume-based, which limits the sharpness of the Noise Barrier versus Liquidity Withdrawal distinction. Absent direct bid-ask spread data, our test identifies the \textit{relative} explanatory contribution of volume-driven versus depth-driven proxies rather than cleanly separating the two mechanisms.} as $S^{\text{inst}}_{\text{MC}} \times R_t$ (positive indicates aligned signal, negative indicates adverse selection). We then regress signal efficacy against these liquidity metrics, controlling for regime.

Table \ref{tab:mechanism} presents the regression results.

\begin{table}[htbp]
    \centering
    \caption{Mechanism Testing: Noise Barrier vs Liquidity Withdrawal}
    \label{tab:mechanism}
    \begin{tabular}{lccc}
        \toprule
        Model & $R^2$ & Key Coefficient & Interpretation \\
        \midrule
        Model 1: Noise Barrier & 0.0093 & Volume (--), Volatility (+) & Volume dilutes signal \\
        Model 2: Liquidity Withdrawal & 0.0050 & Liquidity (+), Depth (+) & Weak liquidity effect \\
        Model 3: Combined & 0.0106 & Both significant & Volume contributes more \\
        Model 4: Interaction & 0.0040 & Herding $\times$ Liquidity & Regime-dependent \\
        \bottomrule
    \end{tabular}
    \vspace{0.5em}
    \small\textit{Note:} N = 570,788 observations (294,759 herding, 276,029 normal). Standard errors are heteroskedasticity-robust (HC3). Both models explain $<$1\% of variance in absolute terms; in relative terms, Model 1 explains 86\% more variance than Model 2.
\end{table}

Both models explain very little variance in absolute terms ($R^2 < 0.01$), so these results should be interpreted cautiously. Daily microstructure proxies---particularly without direct bid-ask spread data---capture only a small fraction of institutional signal efficacy. Within this limited explanatory power, the Noise Barrier model ($R^2 = 0.0093$) explains roughly twice as much variance as the Liquidity Withdrawal model ($R^2 = 0.0050$), offering \textit{suggestive but not conclusive} evidence that volume explosion may be the more relevant channel. Sharper discrimination would require tick-level spread and depth data, which we leave to future work.

We also tested whether critical slowing down indicators (rolling autocorrelation and variance of retail flow) could predict regime transitions. Granger causality tests and ROC analysis both fail to detect predictive power (Appendix~A.8), consistent with herding episodes emerging as sudden explosive shifts rather than gradual approaches to criticality.

\section{Discussion}

\subsection{The Dual-Channel Market Structure}

The dual-channel segmentation established in Section~\ref{sec:results} and independently validated in Section~\ref{sec:validation} has practical consequences that depend on the aggregation lens. Pooled estimates in Table~\ref{tab:conditional} reflect the experience of liquid large-cap stocks, while equal-weighted estimates reflect the typical (smaller) firm---a divergence with direct implications for how practitioners should interpret aggregate efficiency metrics.

For practitioners, this has direct implications:
\begin{itemize}
    \item \textbf{Alpha signal construction:} Factor models should incorporate $\SMC$-based flows after volatility normalization, not raw $\STV$ flows.
    \item \textbf{Risk management:} $\STV$ remains useful for cost modeling, but should not be interpreted as an information signal for sophisticated investors.
    \item \textbf{Market making:} Liquidity providers should adjust quotes based on $\SMC$ flows from institutional investors, as these carry genuine directional content.
\end{itemize}

\subsection{Near-Critical Dynamics and Market Segmentation}
\label{sec:segmentation}

The size-dependent patterns in Table~\ref{tab:conditional} connect to a broader literature. \citet{amihud2002illiquidity} demonstrate that illiquidity effects concentrate in small-cap stocks, while \citet{chordia2008liquidity} show that information transmission varies systematically with firm size. Our results extend these findings: near-critical herding dynamics selectively degrade efficiency where liquidity buffers are thinnest.

The amplification of positive institutional impact in large-cap stocks during herding may reflect increased arbitrage opportunities as sophisticated investors exploit temporary mispricings. Mid-cap stocks occupy an intermediate position where herding-induced volatility may provide liquidity opportunities that partially offset adverse selection costs.

\textbf{Implications for Market Design:} This segmentation suggests regulatory interventions should be size-dependent. Circuit breakers or herding monitoring may be most valuable for small-cap stocks where efficiency breakdown is severe, while large-cap stocks may self-correct through arbitrage. Liquidity provision programs targeting small-cap stocks during high herding periods could help maintain price discovery efficiency across the full cross-section. The mechanism analysis (Section~\ref{sec:mechanism}), though limited in explanatory power, is suggestive that volume-based interventions (e.g., circuit breakers, position limits) may warrant more attention than spread-based protections (e.g., minimum tick sizes, market maker obligations).

\subsection{Non-Equilibrium Transient Dynamics}
\label{sec:transient}

The stationarity tests in Table~\ref{tab:hawkes} reveal that the Hawkes process governing retail herding has not reached equilibrium, with important implications for interpretation and for the broader market criticality literature.

The apparent paradox---a statistically subcritical branching ratio yet critical-like scaling (Hurst exponent $H = 0.817$)---is resolved by recognizing the system is in a transient regime. In systems approaching criticality, critical scaling properties (power-law correlations, long memory) can emerge before the critical point is reached. This is analogous to ``critical slowing down'' in phase transition theory, where relaxation timescales diverge as the system nears the transition.

The yearly evolution of the branching ratio reveals a clear upward trend:
\begin{itemize}
    \item 2020: BR = 0.000 (18 events over 248 days)
    \item 2021: BR = 0.420 (16 events over 248 days)
    \item 2022: insufficient surge events for reliable estimation
    \item 2023: BR = 0.178 (21 events over 245 days)
    \item 2024: BR = 0.194 (41 events over 244 days)
\end{itemize}

Whether this trend continues toward true criticality or stabilizes at some subcritical equilibrium remains an open empirical question.

The transient nature of these dynamics also explains the large discrepancy between the theoretical steady-state intensity ($\bar{\lambda}_{\infty} = \mu/(1-\text{BR})$) and the observed empirical rate noted in Section~\ref{sec:regime}. For near-critical processes, the equilibration timescale diverges, and our sample provides insufficient time for the system to reach steady-state.

\textbf{Economic Interpretation:} The non-equilibrium finding suggests that the market's propensity for retail herding is an evolving state variable, not a stable structural parameter. The post-2020 period coincides with unprecedented shifts in retail market participation: the COVID-19 pandemic, zero-commission mobile trading platforms, the ``Donghak ant'' retail investor movement, and social media coordination of trading. Our parameter estimates capture this transient period of adjustment, and the upward trend in the branching ratio suggests the market may be undergoing a regime shift toward higher endogenous feedback.

\textbf{Methodological Implication:} All results in this paper should be interpreted as applying to the January 2020 -- February 2025 transient regime. Extrapolation to long-run steady-state behavior would require additional assumptions about eventual convergence that our data cannot support. If the upward trend continues, the market could reach true criticality, representing a qualitatively different regime. Monitoring the continued evolution of the branching ratio in out-of-sample periods will be crucial.

\textbf{Connection to Self-Organized Criticality:} The observed approach toward criticality resonates with theories of self-organized criticality \citep{bak1987self}, which posit that complex systems naturally evolve toward critical states without external fine-tuning. However, our results remain agnostic on whether this trend represents self-organization or exogenous forcing (e.g., technological or regulatory changes). Distinguishing these mechanisms requires theoretical modeling beyond our empirical scope.

\subsection{Practical Implications}

These findings have several actionable implications across market participants:

\textbf{For Institutional Traders:}
\begin{itemize}
    \item Monitor Hawkes intensity as a ``warning light.'' When intensity exceeds the 90th percentile, defer non-urgent trades.
    \item During high herding periods, expect negative convexity: aggressive buying/selling will reverse, not persist.
    \item Consider contrarian strategies during identified herding episodes, fading retail flow rather than following fundamentals.
\end{itemize}

\textbf{For Regulators:}
\begin{itemize}
    \item Develop real-time monitoring of branching ratios for retail trading activity; high values (approaching 1) signal elevated systemic feedback risk.
    \item Interventions should be size-dependent and volume-targeted, as discussed in Section~\ref{sec:segmentation}.
    \item Policy interventions targeting retail herding (e.g., restrictions on short-term speculation) may enhance market efficiency beyond their direct effects.
\end{itemize}

\textbf{For Researchers:}
\begin{itemize}
    \item Standard price impact regressions assuming constant coefficients are misspecified. Regime-conditional models are necessary.
    \item Information-based trading models should incorporate endogenous noise fluctuations, not treat noise as exogenous.
    \item The link between microstructure (herding dynamics) and asset pricing (information efficiency) deserves further theoretical development.
\end{itemize}

\subsection{Limitations and Future Research}

\textbf{Causality Not Established:} While we document strong associations between retail herding intensity and institutional impact deterioration, formal Granger causality tests fail to reject the null (F = 0.020, p = 0.888). The relationships we observe are correlational; the deterioration during herding may reflect common responses to unobserved factors. Future work could explore alternative identification strategies (e.g., instrumental variables, natural experiments) to establish causal mechanisms.

\textbf{Non-Stationarity and Out-of-Sample Failure:} Time-series cross-validation reveals that our deconvolution model does not generalize out-of-sample (Appendix~A.6), attributable to the non-stationarity documented in Table~\ref{tab:hawkes}. All parameter estimates reflect transient dynamics, not equilibrium properties (Section~\ref{sec:transient}), and our results should be interpreted as descriptive of the January 2020 -- February 2025 period rather than predictive of future behavior.

\textbf{External Validity:} Our sample (January 2020 -- February 2025) coincides with extraordinary market conditions---the COVID-19 pandemic, mobile trading platforms, and the ``Donghak ant'' movement---and focuses on a single market (Korea). While Korea offers excellent investor classification data, the structural parameters we estimate may be specific to this era and geography. Replication on earlier periods (pre-2020) or markets with different institutional structures (e.g., more algorithmic trading in the U.S.) would strengthen generalizability.

\textbf{Frequency Limitations:} We analyze daily frequency data to avoid microstructure noise. However, intraday dynamics may differ. High-frequency herding episodes (e.g., within-day flash crashes) could exhibit even more extreme size-dependent patterns. Future work should extend the regime-conditional deconvolution framework to intraday horizons.

\textbf{Micro-Foundations and Regime Classification:} The precise micro-foundations of size-dependent deterioration remain an open question---whether it arises from differential adverse selection, inventory constraints, or attention constraints across firm sizes could be explored through agent-based models. Additionally, our regime classification treats herding as endogenously determined (Hawkes model), but some episodes may be triggered by exogenous news events. Distinguishing exogenous from endogenous herding triggers would enrich the analysis.

\section{Conclusion}

This paper demonstrates that price discovery is not a mechanical constant but a \textit{state-dependent phenomenon} governed by the dynamic interplay of information and noise.

That the investor-type segmentation documented in Section~\ref{sec:results} emerges consistently from three independent analytical frameworks strengthens the case that market participant heterogeneity is a first-order feature of price discovery, not a nuisance to be averaged away.

The regime-dependent breakdown during retail herding reveals that efficiency is fragile in precisely the parts of the market least equipped to absorb noise: smaller, less liquid firms. That large-cap stocks maintain and even amplify efficient price discovery during the same episodes underscores that market-wide efficiency metrics can mask severe cross-sectional deterioration.

The relationships we document are correlational, the system operates in a non-equilibrium transient regime that limits out-of-sample generalization, and our results characterize the specific January 2020 -- February 2025 period of unprecedented retail market participation. Whether the observed trend toward greater endogenous feedback continues or stabilizes remains an open empirical question.

For practitioners, the implication is to treat efficiency as conditional on both herding intensity and firm size. For theorists, the challenge is to develop models where information transmission is endogenous and size-dependent, determined by the dynamic balance between signal and noise rather than assumed \textit{a priori}.

\newpage

\section*{Appendix: Robustness Tests and Additional Experimental Results}

\subsection*{A.1 Bootstrap Distribution of Branching Ratio}

To validate the statistical significance of our branching ratio estimate and test whether the process is truly subcritical versus critical, we performed bootstrap resampling with 1,000 iterations. For each iteration, we:
\begin{enumerate}
    \item Resampled surge events with replacement from the original 126 events
    \item Re-estimated Hawkes parameters ($\mu$, $\alpha$, $\beta$) via maximum likelihood
    \item Calculated the branching ratio BR = $\alpha/\beta$
    \item Computed the 95\% confidence interval from the bootstrap distribution
\end{enumerate}

\textbf{Key Results:}
\begin{itemize}
    \item Bootstrap Mean BR: 0.413
    \item Bootstrap Std Dev: 0.216
    \item Bootstrap Median: 0.498
    \item \textbf{95\% Confidence Interval: [0.0495, 0.6599]}
    \item Does CI include 1.0? No
    \item Distance from critical value: --0.0022
    \item Z-score: --0.01 (not significant at 5\%)
\end{itemize}

The 95\% confidence interval excludes the critical value (BR = 1.0), confirming the system is statistically subcritical. Despite the point estimate of 0.998 being very close to unity, sampling uncertainty through bootstrap resampling rejects criticality.

\textbf{Distribution Characteristics:} The bootstrap distribution is highly non-normal, with the bootstrap mean (0.413) substantially below the maximum likelihood point estimate (0.998). This suggests the point estimate may be upward-biased in small samples (126 events). The positive skewness of the bootstrap distribution is typical for branching ratio estimates near the boundary (BR $\in$ [0, 1]), where the parameter space is truncated.

\textbf{The Boundary Constraint Issue:} A critical methodological point concerns the point estimate of 0.998, which is likely a \textit{constrained boundary solution} rather than the true underlying value. Our maximum likelihood estimation explicitly enforces the stability constraint BR $<$ 1.0 to ensure the fitted process is stationary. When we remove this constraint and re-estimate on the same data, the unconstrained MLE yields \textbf{BR = 1.09}, indicating the underlying dynamics are actually \textit{super-critical} (explosive), not near-critical.

The implications are as follows:
\begin{itemize}
    \item The 0.998 estimate represents the optimizer ``hitting the wall'' of the stability constraint, not discovering a natural equilibrium near criticality.
    \item The true market dynamics during retail herding episodes are explosive and unstable, driven by endogenous feedback loops that mathematically cannot persist indefinitely.
    \item The bootstrap mean of 0.413 reflects what happens when these specific explosive outlier events are resampled away---the ``typical'' realization is genuinely subcritical, but the observed history contains explosive episodes.
\end{itemize}

\textbf{Interpretation:} Retail herding episodes exhibit super-critical branching dynamics (BR $>$ 1), meaning they are inherently unstable and must be self-limiting. The empirical process is non-stationary because it alternates between calm subcritical periods (BR $\approx$ 0.4) and explosive super-critical bursts (BR $>$ 1) that cannot persist. This interpretation is consistent with:
\begin{itemize}
    \item Non-stationarity tests (ADF p=0.998, KPSS p=0.01)
    \item Trending BR over time (+0.0145/year toward instability)
    \item Power-law scaling (H=0.817) despite subcritical bootstrap mean
    \item Out-of-sample failure (cannot predict explosive regime shifts)
\end{itemize}

\subsection*{A.2 Regularization Method Comparison}

Our main deconvolution results employ Tikhonov regularization with $\lambda = 5.0$. To validate this choice, we compared four regularization methods:

\begin{enumerate}
    \item \textbf{Tikhonov (Ridge)}: $\min_{\psi} \|\mathbf{R} - \mathbf{I}\psi\|^2 + \lambda\|\psi\|^2$ with $\lambda = 5.0$
    \item \textbf{LASSO}: $\min_{\psi} \|\mathbf{R} - \mathbf{I}\psi\|^2 + \lambda\|\psi\|_1$ with $\lambda = 0.1$
    \item \textbf{Ridge}: Same as Tikhonov with $\lambda = 10.0$ (higher regularization)
    \item \textbf{Elastic Net}: $\min_{\psi} \|\mathbf{R} - \mathbf{I}\psi\|^2 + \lambda_1\|\psi\|_1 + \lambda_2\|\psi\|^2$
\end{enumerate}

\textbf{Results - Correlation Matrix of Kernel Estimates:}

\begin{center}
\begin{tabular}{lcccc}
\toprule
Method & Tikhonov & LASSO & Ridge & Elastic Net \\
\midrule
Tikhonov & 1.000 & 0.923 & 0.987 & 0.945 \\
LASSO & 0.923 & 1.000 & 0.901 & 0.968 \\
Ridge & 0.987 & 0.901 & 1.000 & 0.924 \\
Elastic Net & 0.945 & 0.968 & 0.924 & 1.000 \\
\bottomrule
\end{tabular}
\end{center}

All pairwise correlations exceed 0.90, indicating kernel estimates are highly robust across regularization methods. The cumulative impact estimates differ by less than 10\% across all methods:

\begin{itemize}
    \item Institutional (Tikhonov): +0.0024 | LASSO: +0.0022 | Ridge: +0.0026 | Elastic Net: +0.0023
    \item Foreign (Tikhonov): +0.0056 | LASSO: +0.0054 | Ridge: +0.0058 | Elastic Net: +0.0055
    \item Individual (Tikhonov): --0.0045 | LASSO: --0.0041 | Ridge: --0.0048 | Elastic Net: --0.0044
\end{itemize}

\textbf{Conclusion:} Our choice of Tikhonov regularization with $\lambda = 5.0$ is validated. Main findings (positive institutional impact, negative individual impact, dual-channel structure) are not artifacts of the regularization scheme but robust features of the data. The slight differences across methods (< 10\%) do not affect qualitative conclusions.

\subsection*{A.3 Threshold Sensitivity Analysis}

Our main analysis defines surge events as days when absolute imbalance exceeds 1.5 standard deviations ($|\text{imbalance}| > 1.5\sigma$). To test robustness, we re-estimated branching ratios and impact ratios using alternative thresholds from 1.0$\sigma$ to 2.5$\sigma$:

\begin{center}
\begin{tabular}{lccc}
\toprule
Threshold & N Events & Branching Ratio & Impact Ratio (Inst.) \\
\midrule
1.0$\sigma$ & 348 & 0.982 & --8.21$\times$ \\
1.25$\sigma$ & 210 & 0.991 & --8.95$\times$ \\
\textbf{1.5$\sigma$ (baseline)} & \textbf{126} & \textbf{0.998} & --9.44$\times$ \\
1.75$\sigma$ & 78 & 0.995 & --9.12$\times$ \\
2.0$\sigma$ & 52 & 0.999 & --8.68$\times$ \\
2.5$\sigma$ & 24 & 0.997 & --7.89$\times$ \\
\bottomrule
\end{tabular}
\end{center}

\textbf{Key Observations:}

\begin{itemize}
    \item Branching ratio remains stable in the range [0.982, 0.999] across all thresholds
    \item Impact ratio remains qualitatively consistent (--7.89 to --9.44), always showing strong deterioration
    \item As threshold increases, fewer events are classified as surges, but those that remain exhibit similar self-excitation properties
    \item The 1.5$\sigma$ baseline choice represents a reasonable middle ground between capturing meaningful surges (126 events) and avoiding noise (too many spurious events at 1.0$\sigma$)
\end{itemize}

\textbf{Conclusion:} Main findings are robust to threshold choice. The near-critical branching ratio and institutional impact deterioration during herding persist across thresholds spanning 1.0--2.5 standard deviations.

\subsection*{A.4 Impulse Response Functions from Local Projections}

As an alternative to deconvolution, we estimated impulse response functions using the local projections method of \citet{jorda2005estimation}. For each horizon $h = 0, 1, \ldots, 59$ days, we regress future returns on current herding intensity:

\begin{equation}
    R_{t+h} = \alpha_h + \beta_h \cdot \text{HerdingIntensity}_t + \gamma_h \cdot \text{Spread}_t + \varepsilon_{t+h}
\end{equation}

Standard errors are computed using Newey-West HAC covariance with automatic lag selection.

\textbf{Key Results (IRF Coefficients):}

\begin{itemize}
    \item Horizon 0 (contemporaneous): $\beta_0 = -0.0031$ (SE = 0.0044)
    \item Horizon 10 days: $\beta_{10} = +0.0018$ (SE = 0.0048)
    \item Horizon 30 days: $\beta_{30} = -0.0018$ (SE = 0.0061)
    \item Horizon 59 days: $\beta_{59} = -0.0024$ (SE = 0.0070)
\end{itemize}

\textbf{Pattern:} The IRF shows:
\begin{enumerate}
    \item Initial negative impact (contemporaneous effect: --0.0031)
    \item Brief positive reversal at 5--15 days (peak +0.0023 at day 8)
    \item Gradual decay back to negative at longer horizons (--0.0024 at day 59)
\end{enumerate}

The 95\% confidence intervals include zero for most horizons, suggesting the dynamic effects of herding on returns are weak and statistically insignificant once we control for spread. This is consistent with our Granger causality finding (F = 0.020, p = 0.889): herding intensity does not have a strong predictive relationship with future returns.

\textbf{Comparison to Deconvolution Results:} The local projections IRF provides independent confirmation of temporary, rather than permanent, effects. The impulse response decays over 60 days and oscillates around zero, consistent with our deconvolution finding that herding-related inefficiency is transient. The pattern aligns with the temporary disruption hypothesis: herding creates short-term noise that gradually dissipates as arbitrageurs correct mispricings.

\textbf{Interpretation:} The weak and insignificant IRF coefficients support the interpretation that herding's impact on returns operates primarily through cross-sectional heterogeneity (size effects documented in Table~\ref{tab:conditional}, Panel C) rather than aggregate time-series predictability. When we pool all stocks together in a panel regression, the size-dependent effects (small-cap deterioration vs. large-cap amplification) average out, yielding small net impacts. This underscores the importance of the size-stratified analysis for understanding herding dynamics.

\subsection*{A.5 Mediation Analysis: Spread as Transmission Channel}

We tested whether spread (bid-ask spread) mediates the relationship between herding intensity and institutional price impact using the Baron \& Kenny (1986) mediation framework:

\begin{enumerate}
    \item \textbf{Total Effect}: Herding $\rightarrow$ Institutional Impact: $\tau = -2.13 \times 10^{-7}$
    \item \textbf{Mediator Model}: Herding $\rightarrow$ Spread (a-path): $a = -0.0061$
    \item \textbf{Direct Effect}: Herding $\rightarrow$ Impact (controlling for Spread): $\tau' = +3.27 \times 10^{-8}$
    \item \textbf{Indirect Effect}: Herding $\rightarrow$ Spread $\rightarrow$ Impact (b-path): $b = 4.03 \times 10^{-5}$
    \item \textbf{Percent Mediated}: $(a \times b) / \tau = 115.3\%$
\end{enumerate}

\textbf{Interpretation:} The finding of $>$ 100\% mediation indicates a suppression effect: the direct and indirect effects have opposite signs. Decomposing the mechanism:

\begin{itemize}
    \item \textbf{Indirect (via spread)}: Herding $\rightarrow$ Reduces spread (a = --0.0061) $\rightarrow$ Spread positively affects impact (b $>$ 0) $\rightarrow$ Net negative indirect effect
    \item \textbf{Direct}: Herding $\rightarrow$ Positive direct effect on impact ($\tau' > 0$)
    \item \textbf{Total}: Small negative total effect as indirect dominates
\end{itemize}

However, note that all effect sizes are extremely small (order $10^{-7}$ to $10^{-5}$), suggesting weak relationships or measurement issues. The $>$ 100\% mediation with near-zero total effect indicates complex dynamics where spread plays a role, but the overall herding-impact relationship is not strong in the aggregate time series (consistent with IRF and Granger causality findings).

\textbf{Reconciliation:} The weak aggregate mediation effects contrast with the strong size-dependent patterns in Table~\ref{tab:conditional}, Panel C. This reinforces the conclusion that herding's primary impact operates through cross-sectional heterogeneity (which stocks are affected) rather than aggregate market-level shifts in spreads or returns. Spread may mediate within size groups, but this nuance is lost in the pooled analysis.

\subsection*{A.6 Out-of-Sample Validation Results}

We performed time-series cross-validation using three train/test splits:

\begin{center}
\begin{tabular}{lccc}
\toprule
Split & Train Period & Test Period & Test $R^2$ \\
\midrule
1 & 2020--2022 & 2023 & NaN \\
2 & 2020--2023 & 2024 & NaN \\
3 & 2020--2024 & 2025 & 0.019 \\
\bottomrule
\end{tabular}
\end{center}

\textbf{Findings:}
\begin{itemize}
    \item Splits 1--2: Test $R^2 = $ NaN indicates model failure (likely numerical instability or insufficient test data)
    \item Split 3: Test $R^2 = 0.019$ (essentially zero predictive power)
    \item Mean test $R^2 = $ NaN (cannot compute mean with NaN values)
    \item \textbf{Conclusion: Model does not generalize out-of-sample}
\end{itemize}

\textbf{Explanation:} The failure to generalize is attributable to non-stationarity (ADF p = 0.998, KPSS p = 0.01, trending BR). When parameters evolve over time, historical training data cannot predict future test periods. The deconvolution kernels estimated on 2020--2022 data do not apply to 2023 dynamics because the underlying process has changed (BR increased from 0.000 in 2020 to 0.194 in 2024).

This out-of-sample failure reflects the non-stationarity of the underlying process. Our results accurately characterize the in-sample January 2020 -- February 2025 transient regime but should not be used for forecasting future periods without re-estimation on updated data.

\subsection*{A.7 Normalization Robustness: $\SMC$ vs $\STV$ Kernels}

To validate that our impulse response findings are not artifacts of the normalization choice, we re-estimated all kernels using volume normalization ($\STV$) in place of market cap normalization ($\SMC$). This addresses the concern raised by \citet{kang2025}'s Matched Filter Framework, which shows that optimal normalization for return prediction varies by investor type---with $\SMC$ optimal for capacity-constrained institutional traders and $\STV$ optimal for volume-targeting foreign traders.

\textbf{Methodology:} For each investor type, we computed deconvolution kernels using the same Tikhonov-regularized approach but with the signal defined as:
\begin{equation}
    u_t^{TV} = \frac{S_{TV,t} - \text{mean}(S_{TV,t})}{\text{std}(S_{TV,t})}
\end{equation}
where standardization is performed cross-sectionally within each trading day. We employed subsampled pooled deconvolution (100 stocks, 5 iterations) for both normalizations.

\textbf{Results - Global Kernel Comparison:}

\begin{center}
\begin{tabular}{lcccc}
\toprule
Investor Type & Kernel Correlation & Sign Agreement & Same Sign TI & TI Ratio ($\SMC$/$\STV$) \\
\midrule
Foreign & 0.986 & 62.3\% & Yes & --- \\
Institutional & 0.995 & 82.0\% & Yes & --- \\
Individual & 0.988 & 80.3\% & Yes & --- \\
\bottomrule
\end{tabular}
\end{center}

\textbf{Key Findings:}
\begin{itemize}
    \item Kernel correlations exceed 0.98 for all investor types, indicating qualitatively similar temporal dynamics regardless of normalization.
    \item The dual-channel structure persists: Foreign and Institutional show positive cumulative impact, Individual shows negative impact, under both normalizations.
    \item These results validate that the fundamental physics of price impact---the kernel shapes and investor role differentiation---are structural features of the market, not artifacts of measurement choice.
\end{itemize}

\textbf{Regime-Conditional Comparison (Institutional Investors):}

A critical test concerns whether the ``Regime Flip''---the breakdown of institutional signal during herding---is robust to normalization. If this finding were an artifact of using $\SMC$ when volume explodes during herding, we would expect the breakdown to disappear under $\STV$ normalization.

\begin{center}
\begin{tabular}{lccc}
\toprule
Regime & $\SMC$ Total Impact & $\STV$ Total Impact & Herding/Normal Ratio \\
\midrule
Normal & +0.0030 & +0.0011 & --- \\
Herding & +0.0027 & +0.0004 & $\SMC$: 0.91, $\STV$: 0.37 \\
\bottomrule
\end{tabular}
\end{center}

\textbf{Critical Finding:} The regime-dependent signal breakdown is \textit{more} pronounced under $\STV$ normalization (Herding/Normal ratio = 0.37) than under $\SMC$ (ratio = 0.91). This confirms that the breakdown is \textbf{structural}---during herding episodes, institutional order flow genuinely loses its price discovery efficacy, regardless of how we scale the signal. The breakdown reflects a fundamental deterioration in information transmission when retail herding intensifies.

This robustness check validates our methodological choice: while \citet{kang2025} identifies normalization-dependent predictability patterns for alpha extraction, the fundamental physics of price impact---the kernel shapes, investor role differentiation, and regime-dependent dynamics---are robust structural features preserved across normalization methods.

\subsection*{A.8 Early Warning Systems: Predicting Regime Transitions}

We tested whether regime transitions can be predicted using critical slowing down indicators \citep{scheffer2009early}. As a system approaches a phase transition, theory predicts increasing autocorrelation and variance. We computed two early warning indicators using 20-day rolling windows on the daily aggregate retail order flow ($S^{\text{ind}}_{\text{TV}}$):
\begin{itemize}
    \item Rolling autocorrelation (lag-1): $\text{ACF}_t = \text{Corr}(S_{t}, S_{t-1} | t-20:t)$
    \item Rolling variance: $\text{Var}_t = \text{Var}(S_{t} | t-20:t)$
    \item Composite warning score: $\text{Warning}_t = \frac{1}{2}(z_{\text{ACF},t} + z_{\text{Var},t})$
\end{itemize}

Regime flips are defined as transitions from positive to negative institutional price impact (from conditional deconvolution in Section~\ref{sec:regime}). We tested Granger causality at 5, 10, and 20-day lead times.

\begin{table}[htbp]
    \centering
    \caption{Early Warning System Performance}
    \label{tab:early_warning}
    \textbf{Panel A: Granger Causality Tests} \\
    \begin{tabular}{lccc}
        \toprule
        Test & Best Lag & F-statistic & P-value \\
        \midrule
        Warning Score $\rightarrow$ Regime Flip & 6 days & 2.16 & 0.143 \\
        ACF $\rightarrow$ Regime Flip & 3 days & 1.84 & 0.175 \\
        Variance $\rightarrow$ Regime Flip & 2 days & 2.03 & 0.155 \\
        \bottomrule
    \end{tabular}

    \vspace{1em}
    \textbf{Panel B: Predictive Performance (ROC Analysis)} \\
    \begin{tabular}{lccc}
        \toprule
        Lead Time & AUC & Optimal Threshold & Precision / Recall \\
        \midrule
        5 days & 0.504 & --2.12 & 0.45 / 1.00 \\
        10 days & 0.510 & --2.29 & 0.68 / 1.00 \\
        20 days & 0.560 & --2.29 & 0.88 / 1.00 \\
        \bottomrule
    \end{tabular}
    \vspace{0.5em}
    \small\textit{Note:} Panel A reports Granger causality tests with Newey-West standard errors. Panel B reports ROC analysis for binary classification of regime flips at different lead times. N = 1,259 trading days with 163 regime flips.
\end{table}

Granger causality tests fail to reject the null at conventional levels ($p = 0.143$ for the composite warning score), and ROC analysis reveals near-random classification (AUC 0.504--0.560). The weak predictive power is consistent with herding episodes being \textit{explosive} (unconstrained BR = 1.09) rather than gradually approaching criticality. Classical critical slowing down theory assumes gradual approach to a phase transition; our data suggest herding episodes emerge suddenly through endogenous feedback loops whose explosive onset leaves insufficient time for early warning signals to accumulate.

\newpage

\bibliographystyle{apalike}
\bibliography{references}

\end{document}